\documentstyle[proceedings,epsfig,numreferences]{crckapb}

\begin{opening}
\title{LIQUID--GAS PHASE TRANSITION IN FINITE NUCLEI\protect\\
       WITHIN FERMIONIC MOLECULAR DYNAMICS}

\author{HANS FELDMEIER}
\institute{Gesellschaft f\"ur Schwerionenforschung mbH\\ D-64220 Darmstadt}
\author{J\"URGEN SCHNACK}
\institute{Universit\"at Osnabr\"uck, Fachbereich Physik \\
Barbarastr. 7, D-49069 Osnabr\"uck}
\end{opening}

\runningtitle{LIQUID--GAS PHASE TRANSITION IN FINITE NUCLEI}

\newlength{\CaptionWidth}
\newcommand{\element}[2]{$^{#1}$#2}

\newcommand{\fm}{\mbox{fm}}
\newcommand{\MeV}{\mbox{MeV}}

\newcommand{\Operator}[1]{\raisebox{-1.1ex}{
$\!\!\stackrel{\displaystyle #1}{\sim}$}}

\def\half{\frac{1}{2}\;}
\def\bra#1{\langle \, {#1} \, | \;}
\def\ket#1{\; | \, {#1} \, \rangle}

\newcommand{\braket}[2]{\langle \, {#1} \, | \, {#2} \, \rangle}

\begin{document}

\section{Introduction}

The liquid--gas phase transition of nuclear matter is presently
investigated experimentally in several laboratories \cite{theseproc}.
The task is very difficult because one can manipulate
only finite nuclei and the measured information
on the system is rather indirect.
The difference to macro--physics is not only the smallness
of the system with only about 200 constituents but also that
one cannot control the thermodynamic quantities volume or
pressure. The reason is that one is colliding two nuclei in
order to produce excitation energy and compression. But as there
is no container the system begins to expand into the vacuum
right after the compression and heating phase. Therefore one is
all the time in a transient state where equilibrium in its
original meaning, namely a time--independent stationary macro
state, is not reached.  

The excitation energy of the nuclear system can be deduced by
measuring all energies of the outgoing particles and
clusters. Also the number of nucleons which belong to the
nuclear system under investigation is fairly well known. In
peripheral collisions the so called spectator matter, which is
heated by ablation and participant nucleons which enter the
spectator, moves with a speed close to the beam velocity and can
thus be separated from participant matter. The excited spectator
pieces are assumed to have little compression. 

Central collisions lead to higher
excitations and more compression. Selection of events with high
transverse energies of the outgoing fragments are considered to
be most central and are thus distinguished from more peripheral
collisions. But there is always a certain amount of matter
emitted in the forward backward direction which originates from
the corona.  Due to compression and heating the participant
matter will develop a radial collective flow which obstructs
equilibration. It is however possible to 
estimate its magnitude by assuming local equilibrium and a flow
velocity profile, for example proportional to the distance from
the center of the source. 

A "freeze--out" concept is entering all considerations. Usually
the time interval in which the collisions between the nucleons
and the fragments cease is believed to be short enough so that
global thermodynamic properties like temperature and flow
velocity are frozen in. This allows to infer from the mean
kinetic energies of the fragments the division into
collective and thermal energy. The argument is that the thermal
part of the center of mass motion is proportional to the
temperature and independent on mass while the collective part
is proportional to the mass. Both, measurements and molecular
dynamics calculations support this picture.

Despite all these difficulties the hope is that multifragmention
reactions will give information on the coexistence phase because
at freeze--out several fragments coexist with vapor.
The gas phase should be related to vaporization events which
consist mainly of nucleons and only a few small clusters, while
evaporating compound nuclei should represent the hot liquid. 

The challenge to measure nuclear equations of state has been
accepted not only for astrophysical reasons, like supernova
explosions or neutron stars, but also because the
subject in itself is of interest as one is dealing with a
small charged Fermi liquid which is self--bound by the
strong interaction.

\section{Theoretical Approaches}

Different from experiment a theoretical treatment can impose
thermodynamic conditions like volume and
temperature. Grand canonical mean--field calculations have been
performed since long, both relativistic and
non-relativistic,
e.g. \cite{JMZ84,GKM84,BLV84,SeW86,SCG89}. There are however two
major shortcomings with that. First, a mean--field picture does
not treat the coexistence region properly, fluctuations are
missing and a Maxwell construction is needed. Second, they
cannot describe the experimental non--equilibrium situation so
that a direct comparison with data is not possible. 

In addition there is a general difficulty with canonical or
grand canonical treatments of small systems. In principle
all thermostatic information about a system, including the 
liquid--gas phase transition, is contained in the level density 
$\rho(E,N)$ of the Hamiltonian, where $E$ is the energy and $N$
the particle number. When a phase transition occurs $\rho(E,N)$
shows a rapid increase. In a grand canonical (or canonical) 
ensemble the thermal weight factor is 
$\rho(E,N) \exp \{ -(E-\mu N)/T \}$ 
($T$ temperature, $\mu$ chemical potential) 
so that a sudden increase in 
$\rho(E,N)$ is washed out by the
exponential Boltzmann factor. This insensitivity is annoying for
small systems because there the level density 
$\rho(E,N)$ does not raise so steeply with $E$ or $N$
that the product 
$\rho(E,N) \exp \{ -(E-\mu N)/T \} $  
forms a very narrow peak as a function of $E$ or $N$.  
The micro canonical situation is then preferable as it is
directly sensitive to $\rho(E,N)$ within an interval $\Delta E$
\cite{HuellerGross}. 

Micro canonical statistical models \cite{BDM85,Gro90} are in this
respect well suited but they are static and rely on the assumption
that at freeze--out the system is in global equilibrium, both, in
chemical and kinetic degrees of freedom.

In the following we investigate the liquid--gas phase transition
with a Fermionic Molecular Dynamics simulation. This model can
treat nucleus nucleus collisions as well as equilibrium situations.
We will however concentrate on an experimentally not feasible situation,
namely an excited nucleus which is put in an external field.
This field plays the role of a container so that evaporated nucleons
cannot escape but equilibrate with the remaining nucleus (hot liquid).

\section{Fermionic Molecular Dynamics}
\label{FMDModel}

This section contains a brief outline of Fermionic
Molecular Dynamics (FMD). Details can be found in ref. \cite{FBS95}.
The model describes the many--body system with a parameterized 
antisymmetric many--body state
\begin{eqnarray} 
\ket{Q(t)}=\sum_{all \  P}
sign(P) \ket{q_{P(1)}(t)} \otimes \cdots \otimes \ket{q_{P(A)}(t)}
\end{eqnarray}
composed of single--particle Gaussian wave packets
\begin{eqnarray}
\braket{\vec{x}}{q(t)} &=&
\exp\left\{-\frac{(\,\vec{x}-\vec{b}(t)\,)^2}{2\,a(t)}
    \right\}
\otimes\ket{m_s}\otimes\ket{m_t}
\ , \\
&&
\vec{b}(t) = \vec{r}(t) + i\, a(t)\, \vec{p}(t)
\nonumber
\ ,
\label{gaussian}
\end{eqnarray}
which are localized in phase space at $\vec{r}$ and $\vec{p}$
with a complex width $a$.  Spin and isospin are chosen to be
time--independent in these calculations; they are represented
by their $z$--components $m_s$ and $m_t$, respectively.  Given
the Hamilton operator $\Operator{H}$ the equations of motion for
all parameters are derived from the time--dependent variational
principle (operators are underlined with a tilde)
\begin{eqnarray}
\delta \int_{t_1}^{t_2} \! \! dt \;
\bra{Q(t)}\; i \frac{d}{dt} - \Operator{H}\; \ket{Q(t)} \ &=&\ 0
\ .
\label{var}
\end{eqnarray}
In the present investigation the effective two--body
nucleon--nucleon interaction $\Operator{V}$ in the Hamilton
operator consists of a short--range repulsive and long--range
attractive central potential with spin and isospin admixtures
and includes the Coulomb potential \cite{Sch96}.  
The parameters of the interaction have been
adjusted to minimize deviations between calculated and measured
binding energies for nuclei with mass numbers
$4\le A\le40$.

Besides the kinetic energy $\Operator{T}$
and the nucleon--nucleon interaction $\Operator{V}$
the Hamilton operator $\Operator{H}$ includes an external field 
\begin{eqnarray}
\Operator{U}(\omega)
=
\half m \omega^2 \sum_{i=1}^A \vec{\Operator{x}}_i^2
\end{eqnarray}
which serves as a container.

The container is an important part of the setup because it keeps
the evaporated nucleons (vapor) in the vicinity of the
remaining liquid drop so that it equilibrates with the
surrounding vapor.  The vapor pressure is controlled by the
external parameter $\omega$, which appoints the accessible volume.

In our model the nuclear system is quantal and strongly interacting.
The quantal nature does not allow to deduce the temperature from
the kinetic energy distribution as it is the case for classical
systems with momentum independent forces. The zero--point motion
is always present and does not imply a finite temperature. Due to
the fact that the particles are strongly interacting also a fit
to a Fermi distribution will give wrong answers because even in
the groundstate at zero temperature we have partially occupied
single--particle states. 

Therefore, the concept
of an external thermometer, which is coupled to the nuclear
system, is used in the present investigation. The thermometer
consists of a quantum system of distinguishable particles moving in a
common harmonic oscillator potential different from the
container potential.

The time evolution of the whole system is described by the FMD
equations of motion. For this purpose the many--body trial
state is extended and contains now both, the nucleonic degrees
of freedom and the thermometer degrees of freedom
\begin{eqnarray}
\ket{Q} = \ket{Q_n} \otimes \ket{Q_{th}} \ .
\end{eqnarray}
The total Hamilton operator including the thermometer is given
by
\begin{eqnarray}
\Operator{H} = \Operator{H}_n+ \Operator{H}_{th}
                      + \Operator{H}_{n-th}\ ,\quad\Operator{H}_n=
  \Operator{T} + \Operator{V} + \Operator{U}(\omega)
\end{eqnarray}
with the nuclear Hamilton operator $\Operator{H}_n$
and the thermometer Hamilton operator
\begin{eqnarray}
\Operator{H}_{th} = \sum_{i=1}^{N_{th}} \left(
  \frac{\vec{\Operator{k}}^2(i)}{2\; m_{th}(i)}
+ \half m_{th}(i)\; \omega_{th}^2\; \vec{\Operator{x}}^2(i)
\right) \ .
\end{eqnarray}
The coupling between nucleons and thermometer particles,
$\Operator{H}_{n-th}$, is chosen to be weak, repulsive and of
short range. It has to be as weak as possible in order not to
influence the nuclear system too much. On the other hand it has
to be strong enough to allow for reasonable equilibration
times. Our choice is to put more emphasis on small correlation
energies, smaller than the excitation energy, and to tolerate
long equilibration times.

\begin{figure}[ttt]
\unitlength1mm
\begin{picture}(125,160)
\put(0,120){\epsfig{file=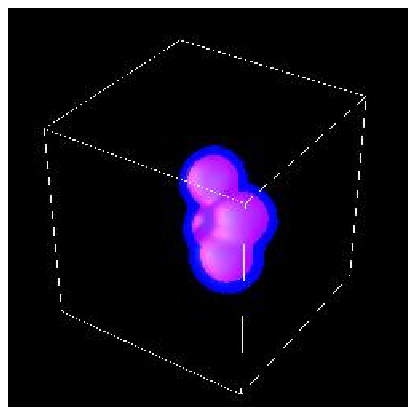,height=40mm}}
\put(0,80) {\epsfig{file=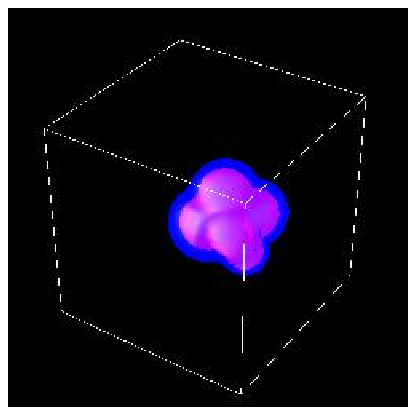,height=40mm}}
\put(0,40) {\epsfig{file=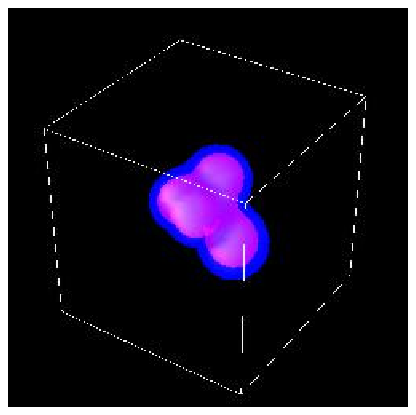,height=40mm}}
\put(0,0)  {\epsfig{file=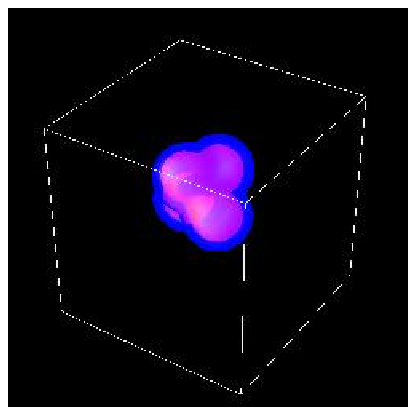,height=40mm}}
\put(42,120){\epsfig{file=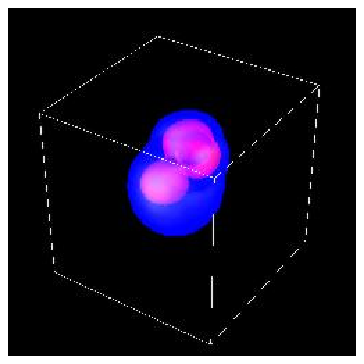,height=40mm}}
\put(42,80) {\epsfig{file=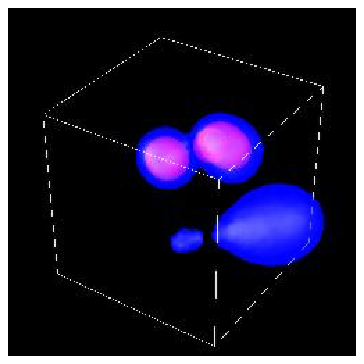,height=40mm}}
\put(42,40) {\epsfig{file=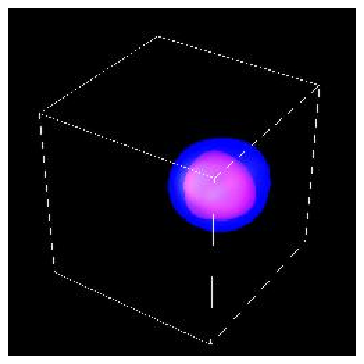,height=40mm}}
\put(42,0)  {\epsfig{file=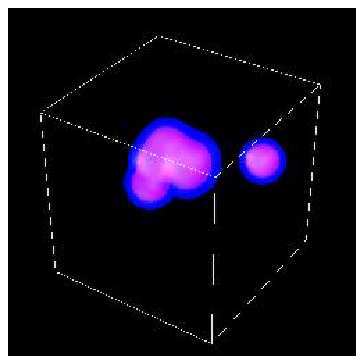,height=40mm}}
\put(84,120){\epsfig{file=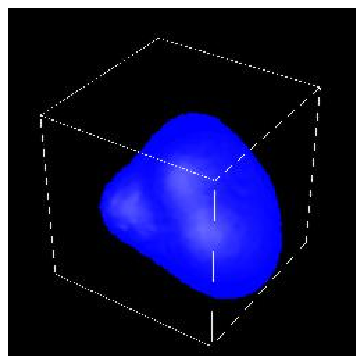,height=40mm}}
\put(84,80) {\epsfig{file=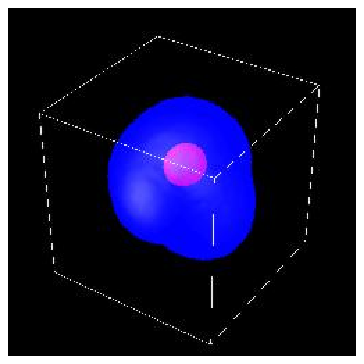,height=40mm}}
\put(84,40) {\epsfig{file=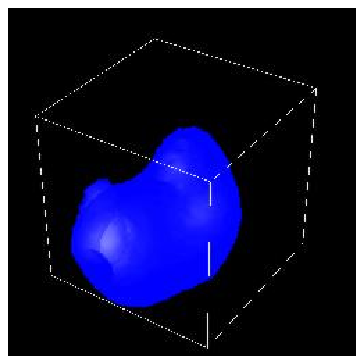,height=40mm}}
\put(84,0)  {\epsfig{file=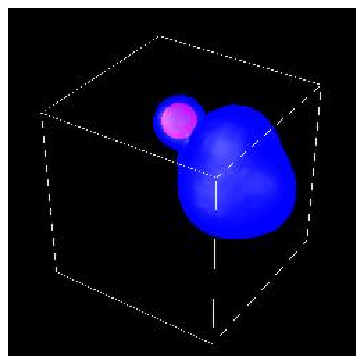,height=40mm}}
\end{picture}
\caption{Snapshots of an excited \element{16}{O} enclosed in a
shallow container potential with $\hbar\omega=1~\MeV$ and 
excitation energy per particle of  3.5~MeV (l.h.s.), 7~MeV
(center) and 11~MeV (r.h.s.) Bright surfaces enclose densities above
$\rho_0/10$ i.e. liquid, darker surfaces
$\rho_0/100$ i.e. gas ($\rho_0$=0.16 fm$^{-3}$ saturation
density).
The cube is 20 fm each side and drawn to help visualizing
three dimensions. }
\end{figure} 
The determination of the caloric curve is done in the following
way. The nucleus is excited by displacing all wave packets from
their ground--state positions randomly. Both, center of mass
momentum and total angular momentum are kept fixed at zero. To
allow a first equilibration between the wave packets of the
nucleus and those of the thermometer the system is evolved over
a long time, about 10000~\fm/c. (A typical time for a nucleon to
cross the hot nucleus is 30 fm/c.) After that a time--averaging of the
energy of the nucleonic system as well as of the thermometer is
performed over a time interval of 10000~\fm/c. During this time
interval the mean of the nucleonic excitation energy
\begin{eqnarray}\label{meanE}
E
&=&
\frac{1}{N_{steps}} \sum_{i=1}^{N_{steps}}
\bra{Q_n(t_i)} \Operator{H}_n \ket{Q_n(t_i)}
\end{eqnarray}
is evaluated. The time--averaged energy
of the thermometer $E_{th}$, which is calculated during the same
time interval, determines the temperature $T$ through the
relation for an ideal gas of distinguishable particles in a
common harmonic oscillator potential (Boltzmann statistics)
\begin{eqnarray}\label{TversusE}
T
&=&
\hbar\omega_{th}
\left[
\ln\left(
\frac{E_{th}/N_{th}+\frac{3}{2}\hbar\omega_{th}}
     {E_{th}/N_{th}-\frac{3}{2}\hbar\omega_{th}}
\right)
\right]^{-1}
\ .
\end{eqnarray}

The general idea behind is the assumption of ergodicity; time
averaging should be equivalent to ensemble averaging. In an
earlier investigation \cite{ScF96} we showed that FMD behaves
ergodically. Time averaged occupation numbers of a weakly
interacting Fermi gas coincided with a Fermi--Dirac
distribution. This however does not mean necessarily that the
system as a
whole is in a grand canonical ensemble because the one--body occupation
numbers represent only a small subset of all degrees of freedom.

We believe that our system is closer to the micro canonical
situation in the sense that the particle number is fixed and a 
pure many--body state $\ket{Q_n(t)}$ is evolved in time. 
This excited state is not an eigenstate of the Hamiltonian but
has a certain width in energy. (If it were an eigenstate it
would be stationary and there would be no dynamical evolution as
seen in Fig~1.)
In principle we could calculate the variance   
$ \bra{Q_n(t)} \Operator{H}_n^2 \ket{Q_n(t)}
-\bra{Q_n(t)} \Operator{H}_n \ket{Q_n(t)}^2 $ 
of the Hamiltonian as a function of time to check our
conjecture. But $\Operator{H}_n^2$ contains a 4--body operator
which means a huge numerical effort.
The coupling to the thermometer also introduces a certain amount
of energy fluctuations but they remain rather small as shown in
the following section.

\section{Caloric Curves}

In Fig. 1 several snap shots of the one--body density 
of a hot nuclear system with 8 neutrons and 8 protons are shown. 
On the left hand side the \element{16}{O} nucleus has been given
an excitation energy per particle of 3.5~MeV by randomly
displacing the wave packets of the ground state.
After equilibration this corresponds to a temperature of about
4~MeV. One sees that the two--body interaction yields an
alpha--particle substructure in $^{16}$O. There is no gas around the
vibrating nucleus because the excitation energy is not high enough to
evaporate particles.

In the center column of Fig.~1 the excitation energy is 7~$A$MeV. 
Bright areas which indicate the liquid are surrounded 
by a cloud of gas (for details see figure caption). 
More over, the nuclear system very often
falls apart into several smaller drops which are embedded in
vapor. 

The right hand side displays the same system but for an
excitation energy of 11~$A$MeV. Here half of the time
no high density areas are visible (first and third frame) and if
a drop is formed it is rather small.

As we shall see later, the two excitation energies
7 and 11 $A$MeV correspond both to  a
temperature around 6 MeV in the coexistence region. It is quite
obvious that the additional excitation energy of 4 MeV per
particle is used to transform liquid to vapor so that we see
a clear first order liquid--gas phase transition. This is
remarkable as we are dealing with only 16 nucleons and the 
dynamical model evolves in time a pure state with a very limited
number of degrees of freedom, actually only eight per particle,
three for mean position, three for mean momentum and two for the 
width. Furthermore, we have a fermion system in which the level
density due to antisymmetrization is much smaller than in
classical mechanics.   

The container is very wide so that the vapor pressure is rather
small. Estimates yield 10$^{-4}$ to 10$^{-2}$~MeV/fm$^3$ which
should be compared to a critical pressure of about 
0.5~MeV/fm$^3$. The container potential itself is at the surface
of the indicated cubes only 1.2 MeV higher than in the center.

To quantify the relation between energy, temperature and
container size we display in Fig. 2 the caloric curve for 
the external parameter $\hbar\omega$ = 1, 6 and 18 MeV, which 
controls the thermodynamic properties of the nucleonic system in
a similar way as the volume. A pronounced plateau is seen in the
plot on the left hand side, 
where the oscillator does not influence the self--bound nucleus
very much. In the middle part the more narrow container
potential is already squeezing the ground state, its energy goes
up to $E/A\approx -5~\MeV$. The plateau is shifted to $T\approx
7~\MeV$ and the latent heat is decreased. On the right hand
side, for $\hbar\omega=18~\MeV$, the coexistence region has
almost vanished and the critical
temperature $T_c$ is reached. 

The solid line represents the relation between temperature and
energy for an ideal Fermi gas in a harmonic oscillator potential
with 
$\omega_{\mbox{\tiny{eff}}}=(\omega^2+\omega_0^2)^{1/2}$,
where 
$\hbar \omega_0=10$ MeV corresponds to the selfconsistent
mean--field of $^{16}$O. The energy zero--point is shifted so
that the ground state of the oscillator is at the FMD
value. The dashed line shows the relation for the external
container, also with the ground state shifted, because even in
the gas phase the particles still feel attraction. Despite the
strong interaction the liquid and the gas phase follow
approximately the picture of an ideal gas in a mean--field. The
coexistence region cannot be approximated by a mean--field
picture like the liquid in a selfconsistent
potential or the gas in the external field.

The "error bars" in temperature and energy represent r.m.s.
deviations from the time averaged mean. There is always an
exchange of energy between thermometer and nuclear system. But
the fluctuations remain rather small. The temperature
fluctuations, which through relation (\ref{TversusE}) are
actually fluctuations in the energy of the thermometer
particles, are larger because the thermometer has a smaller
heat capacity than the nucleons.

\begin{figure}
\unitlength1mm
\begin{picture}(125,46)
\put(0,0){\epsfig{file=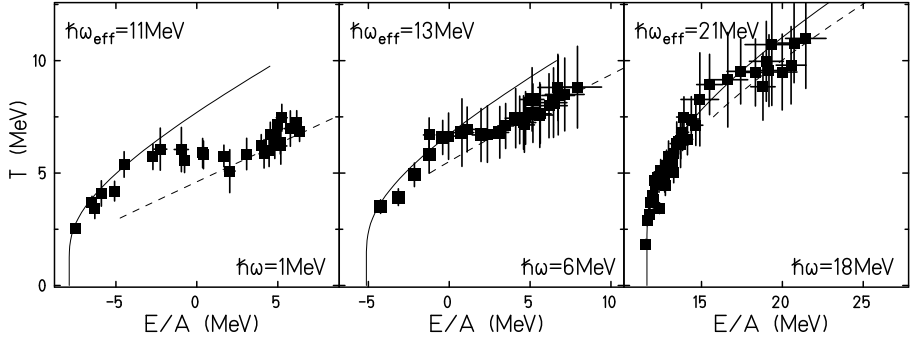,height=45mm}}
\end{picture}
\caption{Caloric curve of \element{16}{O} for the
frequencies $\hbar\omega=1, 6, 18~\MeV$ of the container
potential.  The solid lines show the low temperature behaviour
of an ideal gas of 16 fermions in a common harmonic oscillator
with level spacing $\hbar\omega_{\mbox{\tiny{eff}}}$, the dashed
lines denote their high temperature behaviour in the confining
oscillator ($\hbar\omega$).}
\end{figure} 

The critical temperature $T_c$ can only be estimated from the
disappearance of the coexistence phase in Fig.~2 because the
fluctuations in $T$ and $E$ are rather large. Its value is
about $10~\MeV$ and has to be compared to the results of ref.
\cite{JMZ84,BLV84,CaY96} for finite nuclei including
Coulomb and surface effects. All authors report a week
dependence of the critical temperature on the mass number in the
region from calcium to lead. The result of Jaqaman et al. with
the Skyrme ZR3 interaction \cite{JMZ84} can be extrapolated to
\element{16}{O} to give $T_c\approx8~\MeV$, Bonche et al. \cite{BLV84}
arrive at the same number using the SKM interaction, but got
$T_c\approx11~\MeV$ with the SIII interaction. Close to the last
result is the value extrapolated from ref. \cite{CaY96} where
$T_c\approx11.5~\MeV$ for Gogny's D1 interaction.

\begin{figure}[hhht]
\unitlength1mm
\begin{picture}(125,55)
\put(0,0){\epsfig{file=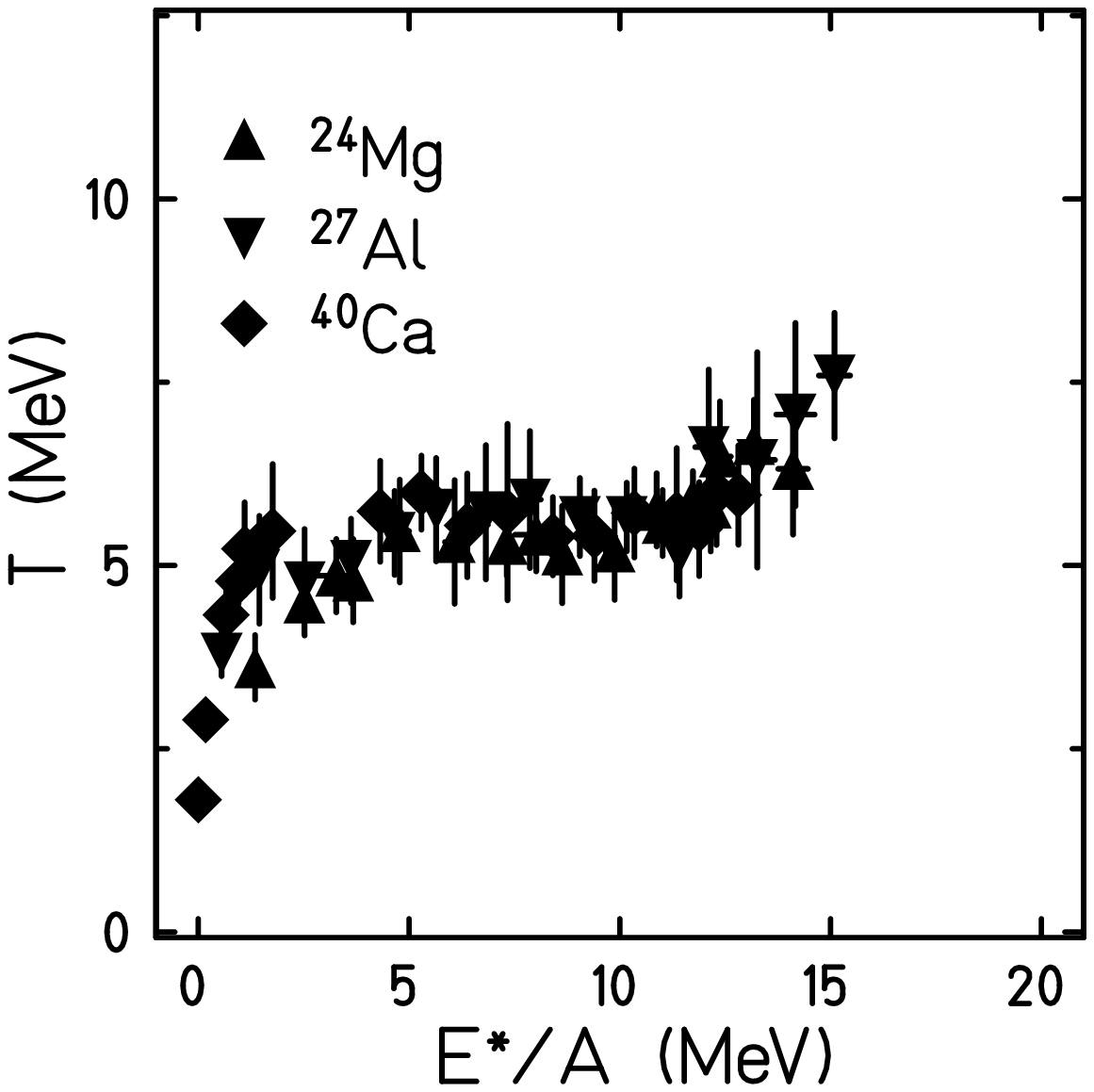,height=50mm,width=46mm}}
\put(70,0){\epsfig{file=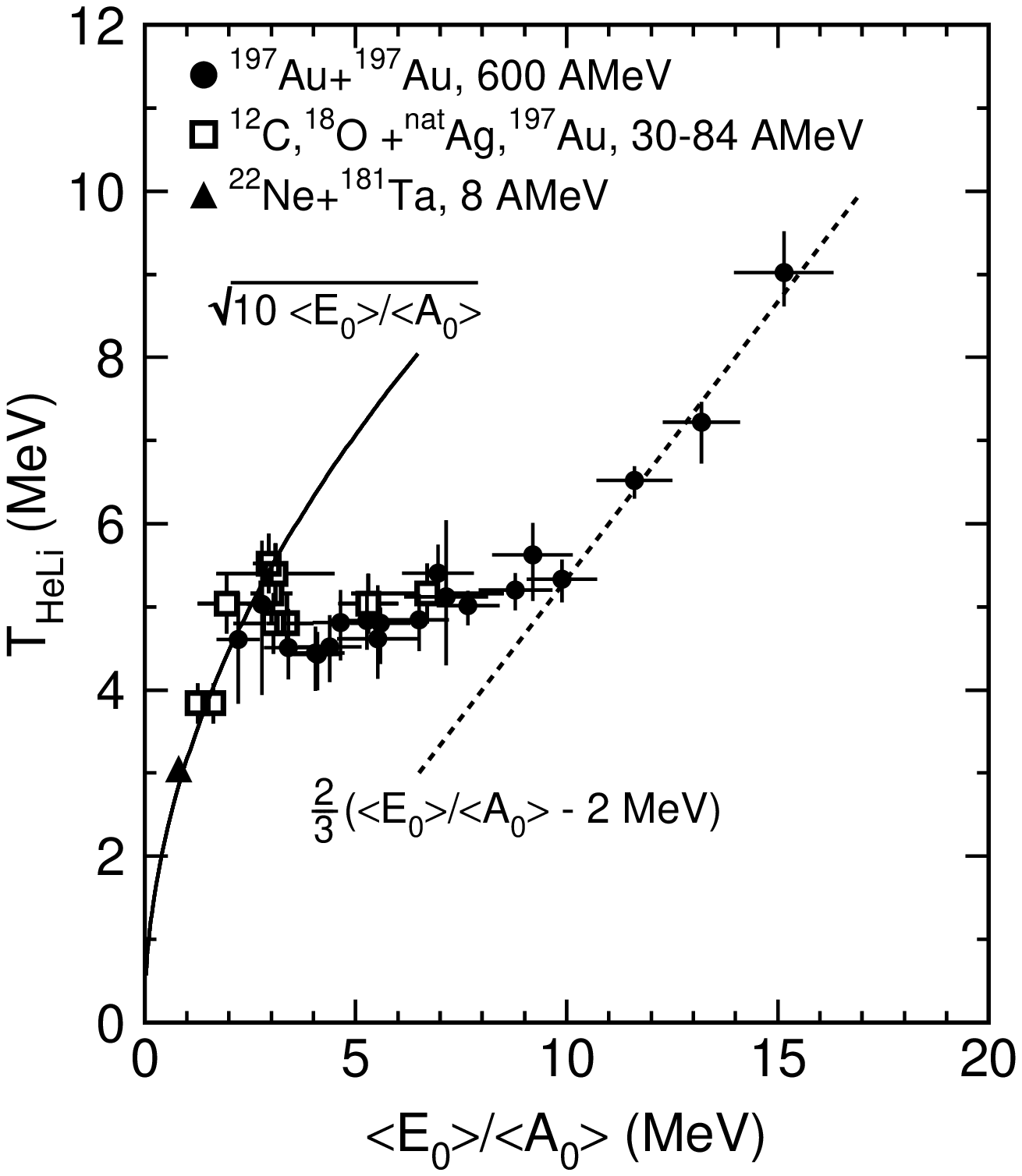,height=50mm}}
\end{picture}
\caption{L.h.s.: caloric curve of \element{24}{Mg}, 
\element{27}{Al} and \element{40}{Ca} at $\hbar\omega=1~\MeV$,
r.h.s.: caloric curve determined by the Aladin group 
from the decay of spectator nuclei.}
\end{figure} 

We determined the relation between the excitation energy and the
temperature also for the larger nuclei 
\element{24}{Mg}, \element{27}{Al} and
\element{40}{Ca} using the same container potential with
$\hbar\omega=1~\MeV$ and summarize them on the left hand side of
Fig. 3. In order to put them on the same scale we
subtract from the averaged energy, defined in eq. (\ref{meanE}), the
respective ground state energies and show the temperature as a
function of excitation energy $E^*$.

Like for \element{16}{O} all caloric curves clearly exhibit three different
parts. Beginning at small excitation energies the temperature
rises steeply with increasing energy as expected for the shell
model. The nucleons remain bound in the excited nucleus which
behaves like a drop of liquid.  At an excitation energy of
$3~\MeV$ per nucleon the curve flattens and stays almost
constant up to about $11~\MeV$. This coexistence plateau at
$T\approx$ 5 to 6 MeV reaches from $E^*/A\approx3~\MeV$ to
about $E^*/A\approx11~\MeV$ where all
nucleons are unbound and the system has reached the vapor
phase. The latent heat at pressure close to zero is hence about
8~MeV. 

One has to keep in mind that the plateau, which due to finite
size effects is rounded, is not the result of a Maxwell
construction as in nuclear matter calculations.  In the
excitation energy range between 3 and $11~\MeV$ per particle an
increasing number of nucleons is found in the vapor phase
outside the liquid phase.  This has been shown in Fig. 1.

The caloric curve shown in Fig. 3 has a striking
similarity with the caloric curve determined by the ALADIN group
\cite{Poc95} which is displayed in the same figure. The position
and the extension of the plateau agree with the FMD calculation
using a containing oscillator potential of
$\hbar\omega=1~\MeV$. Nevertheless, there are important
differences. The measurement addresses an expanding
non--equilibrium system, but the calculation deals with a
contained equilibrium system. In addition the used thermometers
differ; the experiment employs an isotope thermometer based on
chemical equilibrium and the calculation uses an ideal gas
thermometer.  One explanation why the thermodynamic description
of the experimental situation works and compares nicely to the
equilibrium result might be, that the excited spectator matter
equilibrates faster into the coexistence region \cite{FeS97a}
than it expands and cools. The assumption of such a transient
equilibrium situation \cite{PaN95,BDM85,Gro90} seems to work
rather well at least in the plateau region.

\end{document}